\documentclass[aps,superscriptaddress,preprint]{revtex4}

\usepackage{graphicx}
\usepackage{float}
\usepackage[small,hang,bf]{caption}
\usepackage{subfigure}
\usepackage{bm}    
\usepackage{amsmath}  
\usepackage{amssymb}
\usepackage{times}
\usepackage{color} 
\usepackage{tabularx}

\begin{document}

\title{Fast density-matrix based partitioning of the energy over the atoms in a molecule consistent with the Hirshfeld-I partitioning of the electron density}
\author{Diederik Vanfleteren} 
\email{diederik.vanfleteren@ugent.be}
\affiliation{Members of the Ghent Brussels Quantum Chemistry and Molecular Modeling alliance.}
\affiliation{Ghent University, Center for Molecular Modeling, 
Technologiepark 903, B-9052 Zwijnaarde, Belgium}
\author{Dieter Ghillemijn}
\affiliation{Members of the Ghent Brussels Quantum Chemistry and Molecular Modeling alliance.}
\affiliation{Ghent University, Department of Inorganic and Physical Chemistry, Krijgslaan 281 (S3), B-9000 Gent, Belgium}
\author{Dimitri Van Neck}
\affiliation{Members of the Ghent Brussels Quantum Chemistry and Molecular Modeling alliance.}
\affiliation{Ghent University, Center for Molecular Modeling, 
Technologiepark 903, B-9052 Zwijnaarde, Belgium}
\author{Patrick Bultinck}
\affiliation{Members of the Ghent Brussels Quantum Chemistry and Molecular Modeling alliance.}
\affiliation{Ghent University, Department of Inorganic and Physical Chemistry, Krijgslaan 281 (S3), B-9000 Gent, Belgium}
\author{Michel Waroquier}
\affiliation{Members of the Ghent Brussels Quantum Chemistry and Molecular Modeling alliance.}
\affiliation{Ghent University, Center for Molecular Modeling, 
Technologiepark 903, B-9052 Zwijnaarde, Belgium}
\author{Paul W. Ayers}
\affiliation{McMaster University, Department of Chemistry, Hamilton, Ontario L8S 4M1, Canada}

\begin{abstract} 
For the Hirshfeld-I atom-in-molecule model, associated single-atom energies and interaction energies at the Hartree-Fock level are determined efficiently in one-electron Hilbert space. In contrast to most other approaches, the energy terms are fully consistent with the partitioning of the underlying one-electron density matrix. Starting from the Hirshfeld-I atom-in-molecule model for the electron density, the molecular one-electron density matrix is partitioned with a previously introduced double-atom scheme [Vanfleteren D. et al., J Chem Phys 2010, 132, 164111]. Single-atom density matrices are constructed from the atomic and bond contributions of the double-atom scheme. Since the Hartree-Fock energy can be expressed solely in terms of the one-electron density matrix, the partitioning of the latter over the atoms in the molecule leads naturally to a corresponding partitioning of the Hartree-Fock energy. When the size of the molecule or the molecular basis set does not grow too large, the method shows considerable computational advantages compared to other approaches that require cumbersome numerical integration of the molecular energy integrals weighted by atomic weight functions.
\end{abstract}


\maketitle

\section{Introduction}

In the last decades, several algorithms have been developed to identify the atom in the molecule (AIM). For example the Mulliken method \cite{mulliken1955} relies on the attachment of basis functions to atomic centers; Natural Population Analysis \cite{reed1985} relies on the analysis of blocks of the one-electron density matrix (1DM) expressed in some molecular orbital basis; some other methods rely on the partitioning of the molecular electron density in AIM parts. However, not all AIM properties can be directly expressed in terms of the electron density. Of course the density determines all the properties and there are even explicit formulas for them, but these formulas are often computationally impractical. A common example is the kinetic energy of an AIM, which is directly computable from the full 1DM $\rho \left( {{\bf{r}},{\bf{r}}'} \right)$, but not (trivially) from the electron density, the diagonal element of the 1DM. 

A widely adopted solution, which is also the most common recipe employed within the Quantum Theory of Atoms in Molecules (QTAIM) \cite{bader1994}, is to partition such molecular properties using the same atomic weight functions $w_A \left( {\bf{r}} \right)$ that are used to partition the molecular density.  QTAIM uses a zero flux condition of the electron density to define interatomic surfaces, resulting in nonoverlapping atomic regions. Atomic energies are obtained by partitioning the kinetic energy over the atomic domains and using the virial ratio to construct the corresponding atomic potential energy.  Other approaches also rely on atomic regions to partition the molecular energy, but use them in a more general way to decompose the energy into one- and two-atom terms \cite{mayer2001,salvador2001,pendas2004,pendas2005}. Recently, Mandado et al. \cite{mandado2006} partitioned the Hartree-Fock energy in terms of the overlapping Hirshfeld atoms. Their scheme appears useful to investigate proton acidity, the anomeric effect and group transferability, it has also been used in studies of bonding and polarizability \cite{mandado2007,krishtal2008}. However, partitioning of the molecular energy in atomic fragments can be ambiguous as it often relies on the introduction of an arbitrary number of partitionings of unity \cite{mayer2005} into the energy expressions. Moreover, the exact place where the partition of unity is introduced in an expectation value expression can have an important influence on the resulting AIM condensed values \cite{bultinck2007sf}. This ambiguity is often circumvented by the convention to introduce the unity and its partitioning into weight functions ($1=\sum_{A}w_{A}(\bm{r})$) before any operator \cite{mayer2005}. 

To avoid these problems, we partition the Hartree-Fock energy starting from a partitioning of Hartree-Fock molecular 1DM. At the Hartree-Fock level of theory the energy can be expressed solely and directly in terms of a (partitioned) molecular 1DM. A partitioning of the Hartree-Fock molecular 1DM therefore directly leads to a natural partitioning of the energy. In a previous paper \cite{vanfleteren2010} we introduced a double-atom partitioning scheme for the molecular 1DM that is consistent with existing partitioning schemes for the molecular density. In the current work we use molecular 1DM fragments from that scheme to calculate the energy terms naturally associated with these fragments. 
The strategy of this work is therefore to calculate "properties of the molecular fragments", $\int\limits_{} {d{\bf{r}}\left. {{\rm{\hat A}}\rho _A \left( {{\bf{r}},{\bf{r}}'} \right)} \right|_{{\bf{r}} = {\bf{r}}'} }$, instead of "fragmenting molecular properties", $\int\limits_{} {d{\bf{r}}\left. {w_A \left( {\bf{r}} \right){\rm{\hat A}}\rho \left( {{\bf{r}},{\bf{r}}'} \right)} \right|_{{\bf{r}} = {\bf{r}}'} }$. This is in line with the work of Bultinck et al.\cite{bultinck2007sf} for the derivative of the density function with respect to a change in the number of electrons. We also examine the correspondences and differences between both approaches.

From a practical point of view, a fragmentation of molecular properties requires that the expectation value integrals are computed numerically on a large spatial grid. In general there is no simple analytical expression for the atomic weight functions and these weight functions are often not well expressed in one-electron Hilbert space. In contrast, molecular 1DM fragments show a satisfying basis set convergence in one-electron Hilbert space \cite{vanfleteren2010}. Therefore, it is tempting to calculate the atomic and interaction energies of the 1DM fragments in one-electron Hilbert space, avoiding cumbersome numerical integrations in $\bm{r}$-space. As a consequence, we expect to find that significant computational advantages are a major asset of our current approach. 

\section{Atomic density matrices that are consistent with the Hirshfeld-I partitioning of the electron density}
\label{single-atom}

We use the notation $\bm{x}=\bm{r}\sigma$ to specify the single-electron states in coordinate space, where $\sigma$ represents the spin degrees of freedom. The one-electron density matrix (1DM) for an $N$-electron molecule with  wave function $\Psi (\bm{x}_1 ,\dots ,\bm{x}_N )$ is defined as 
\begin{equation}
\rho (\bm{x},\bm{x}') = N \int d\bm{x}_2 \dots \int d\bm{x}_N 
\Psi^\dagger (\bm{x} ,\bm{x}_2 , \dots ,\bm{x}_N )\Psi (\bm{x}' ,\bm{x}_2 , \dots ,\bm{x}_N ).
\end{equation}
We restrict ourselves to molecules with a singlet ground state. In that case 
\begin{equation}
\rho (\bm{x},\bm{x}') = \frac{1}{2}\delta_{\sigma , \sigma'}\rho (\bm{r}, \bm{r}') , 
\label{singlet}
\end{equation}
and the electron spin can be discarded. 
In a previous paper we introduced a double-atom 
partitioning scheme for the molecular spin-summed 1DM \cite{vanfleteren2010}, 
\begin{equation}
\rho (\bm{r}, \bm{r'}) = \sum_{AB} \rho_{AB}(\bm{r}, \bm{r'}),  
\label{eq1}
\end{equation}
in terms of atomic ($A = B$) and diatomic contributions ($A\neq B$), where A and B label atoms. 
The individual contributions are defined as 
\begin{equation}
\rho_{AB}(\bm{r}, \bm{r'}) = \frac{1}{2}\left[w_A(\bm{r})w_B(\bm{r'})
+w_B(\bm{r})w_A(\bm{r'})\right]\rho (\bm{r}, \bm{r'}),    
\label{eq2}
\end{equation}
with positive local weight functions $w_A (\bm{r})$ obeying 
\begin{eqnarray}
\sum_A w_A (\bm{r}) = 1.
\label{sumweights}
\end{eqnarray} 

At the Hartree-Fock level the energy is expressed solely in terms of the 1DM. Therefore, the partitioning of the latter over the atoms in the molecule leads naturally to a corresponding partitioning of the Hartree-Fock energy. However, for a simple energy partitioning we require that (1) it is based on single-atom density matrices (rather than double-atom density matrices) and (2) the electron density of the atoms is local and positive definite. Note that any single-atom density matrix $\rho_{A}(\bm{r}, \bm{r'})$ unavoidably has bad localization properties (in contrast to the double-atom density matrices) \cite{vanfleteren2010}, and therefore we restrict the requirement of locality and positive definiteness to the electron density, the diagonal element of the 1DM. The latter requirement is needed to prevent the energy components to become unrealistically large on the chemical energy scale.
When the single-atom density matrices are defined as
\begin{eqnarray}
\rho_A (\bm{r},\bm{r}') &=& \sum_B \rho_{AB}(\bm{r},\bm{r}'), 
\end{eqnarray}
it is clear that on the diagonal ($\bm{r}=\bm{r}'$) the following definition is obtained,
\begin{equation}
\rho_A (\bm{r}) =\sum_B \rho_{AB}(\bm{r},\bm{r})= w_A (\bm{r})\rho(\bm{r}) 
\label{eq3}
\end{equation}
that is familiar from existing partitioning schemes for the electron density (with good localization properties). Also the fundamental property
\begin{equation}
\sum_A \rho_A (\bm{r},\bm{r}') = \rho(\bm{r},\bm{r}') 
\label{rhoA}
\end{equation}  
is obeyed as a trivial consequence of Eq.~(\ref{eq1}). Note that these single atom density matrices may not be N-representable, an issue that, however, does not stand in the way of obtaining integrated quantities such as energy contributions. 

In order to define a computationally efficient energy partitioning scheme, we assume real wavefunctions and express the partitioned density matrices in the finite basis set used for the molecular calculation.  E.g., in the molecular Hartree-Fock basis set, the following spatial integrals 
\begin{eqnarray}
S^A_{ij}&=& \int d \bm{r} \phi_i (\bm{r}) w_A (\bm{r})\phi_j(\bm{r}),
\end{eqnarray}
that represent elements of the regular atomic overlap matrix (AOM), are sufficient to determine the single-atom 1DM in this basis, 
\begin{equation}
(\rho_A)_{ij}= \int d \bm{r} d \bm{r'} \phi_i (\bm{r})\rho_A (\bm{r},\bm{r}')\phi_j(\bm{r}').  
\end{equation}
The precise expressions for the density matrices are:
\begin{eqnarray}
(\rho_A)_{ij} &=& \sum_n d_n \frac{1}{2}\sum_B (S^A_{in}S^B_{jn}+S^B_{in}S^A_{jn}) = \frac{1}{2}(d_{i}+d_{j}) S^A_{ij},
\label{dens1}
\end{eqnarray}
where a simple sum rule 
\begin{eqnarray}
\sum_B S^B_{jn} = \delta_{jn}
\label{sumrule}
\end{eqnarray}
is used to simplify the expression in Eq. (\ref{dens1}).

In order to ensure that the single-atom density $\rho_{A}(\bm{r})$ is local, it can be made to coincide with the AIM density of some well-established density partitioning scheme like Hirshfeld \cite{hirshfeld1977}, iterative Hirshfeld \cite{bultinck20071}, Iterated Stockholder Atoms \cite{lillestolen2008,bultinck2009} or QTAIM \cite{bader1991,bader1994,popelier2000},
\begin{equation}    
\rho^{AIM}_A (\bm{r}) = h_A(\bm{r}) \rho(\bm{r})
\label{eq5}
\end{equation}
in which the AIM density is obtained by multiplying the molecular density 
$\rho(\bm{r})$ with the characteristic weight function for that AIM technique $h_A (\bm{r})$. 
One simply takes 
\begin{equation}
w_A(\bm{r})\equiv h_A (\bm{r}).
\label{eq6}
\end{equation}
An alternative to Eq. (\ref{eq3}), consisting of distributing the diatomic contributions in a weighted manner, was also investigated in \cite{vanfleteren2010}, but will be discarded here because it gave inferior results.

\section{Energy decomposition  \label{sec-en-decom}}
\label{energy decomposition}

\subsection{Self-energies and interaction energies of the 1DM fragments}

Following the ideas behind the Interacting Quantum Atoms (IQA) \cite{blanco2005,pendas2006a,pendas2006b}, but from the perspective of density matrix fragments in one-electron Hilbert space, each single-atom 1DM with elements $(\rho_A)_{ij}$ constructed in section \ref{single-atom} can be thought of as forming an atomic subsystem \cite{pendas2007} when combined with the nucleus on center A. The molecular energy can then be decomposed as the sum of the ``self-energies'' of the atomic subsystems and the interaction energies between them:
\begin{eqnarray}
E^{mol}=\sum_{A} E^{self}_{A} + \sum_{A < B} E^{int}_{AB}.
\label{reconstruction}
\end{eqnarray}
At the closed-shell Hartree-Fock level, the self-energy of such an atomic subsystem corresponds to:
\begin{eqnarray}
E^{self}_{A}&=& \sum_{ij} (t+V_{A}^{ext})_{ij} (\rho_{A})_{ij}  + \frac{1}{2}\sum_{ijkl} V_{ijkl}\left( (\rho_{A})_{ik}(\rho_{A})_{jl} - \frac{1}{2} (\rho_{A})_{il}(\rho_{A})_{jk} \right) 
\label{selfenergy}
\end{eqnarray}
where
\begin{eqnarray}
t_{ij}&=&\int \phi_{i}^{}(\bm{r}) \hat{t} \phi_{j}(\bm{r})d\bm{r} \nonumber \\
(V_{A}^{ext})_{ij}&=&\int \phi_{i}^{}(\bm{r}) \frac{Z_{A}}{|\bm{r}-\bm{R}_{A}|} \phi_{j}(\bm{r})d\bm{r} \nonumber \\
V_{ijkl}&=&\int \phi_{i}^{}(\bm{r}_{1})\phi_{j}^{}(\bm{r}_{2}) \frac{1}{|\bm{r}_{1}-\bm{r}_{2}|} \phi_{k}(\bm{r}_{1})\phi_{l}(\bm{r}_{2})d\bm{r_{1}}d\bm{r_{2}} \nonumber \\
\label{definitions}
\end{eqnarray}
In Eq. (\ref{definitions}), $\hat{t}$ is the kinetic energy operator and $Z_{A}$ is the nuclear charge on an atom A with nuclear coordinate $\bm{R}_{A}$. $V_{ijkl}$ are two-electron integrals that are not anti-symmetrized. Note that it is implied in Eq. (\ref{selfenergy}) that the atomic subsystems are spin-averaged, consistent with Eq.(\ref{singlet}-\ref{eq2}).
The atomic subsystems interact with each other, according to the following expressions:
\begin{eqnarray}
E^{int}_{AB}&=& \frac{Z_{A}Z_{B}}{r_{AB}} + \sum_{ij}(V_{A}^{ext})_{ij}(\rho_{B})_{ij} + \sum_{ij}(V_{B}^{ext})_{ij}(\rho_{A})_{ij} \nonumber \\ && + \sum_{ijkl} V_{ijkl}\left( (\rho_{A})_{ik}(\rho_{B})_{jl} - \frac{1}{2} (\rho_{A})_{il}(\rho_{B})_{jk} \right) \label{subinter} \\ 
E^{int}&=& \sum_{A<B} E^{int}_{AB}.
\end{eqnarray}
$E^{int}$ represents the total interaction energy of all subsystems within the molecule. $E^{int}_{AB}$ quantifies all interactions between atom pairs in molecules, including the interaction between atom pairs that do not share a chemical bond. This quantity is also useful to assess the strength of the interactions in a ring structure. $E^{int}_{AB}$ does not depend on the choice of reference atoms, it only depends on the AIM. The localized character of the Hirshfeld-I densities ensures that the promotion and interaction energies are within a reasonable range of values, although they are often significantly larger (in absolute value) than typical ``bond energies''. 

\subsection{Promotion energies}

Since it is well known that the molecular environment induces only relatively small changes in atomic energy \cite{pendas2007}, atomic energies are usually referenced to the energy of the isolated atoms. Atoms within a molecular environment are slightly distorted compared to the isolated atoms. This implies that their energy, with respect to the Hamiltonian of the isolated atoms, has increased. This energy increase is called the atomic promotion energy $E^{prom}_{A}$. When the atom in a molecule is identified with the atomic subsystem defined above, its promotion energy is the difference between the (Hartree-Fock) atomic self-energy $E^{self}_{A}$ in Eq. (\ref{selfenergy}) and the Hartree-Fock energy $E_{A}^{0}$ of the isolated atom:
\begin{eqnarray}
E^{prom}_{A}&=& E^{self}_{A} - E_{A}^{0} \\
E^{prom}&=& \sum_{A} E^{prom}_{A}.
\end{eqnarray}

The atomic promotion energies can be considered to result from three successive processes: first a charge transfer step that accounts for the fact that the AIM has a different atomic charge from the isolated atom.
\begin{eqnarray}
\Delta E_{A}^{CT} = E_{A}^{0}(Q_{A})- E_{A}^{0}(0)
\end{eqnarray}
Classically, the reference for this step is the neutral isolated atom in its ground state $E_{A}^{0}(0)$.
$E_{A}^{0}(Q_{A})$ represents the charged isolated atom, and is approximated as the linearly interpolated value between the energies of the isolated atoms with an integer number of electrons $N<(Q_{A}=N+a)<N+1 $. In principle, the HF energy as a function of the number of electrons N is a concave curve between the integers \cite{cohen2003,mori-sanchez2006,mandado2006,perdew2007,cohen2008}. The assumption that it is linear is based on the fact that this holds for the exact energies \cite{perdew1982,yang2000,ayers2008}. This leads to $E_{A}^{0}(Q_{A})$ computed as:
\begin{eqnarray}
E_{A}^{0}(Q_{A}) = a E_{A}^{0}(N+1) + (1-a) E_{A}^{0}(N).
\end{eqnarray}

The interpolated state $E_{A}^{0}(Q_{A})$ is characterized by a value of $\langle S^{2}\rangle$ (where S is the spin angular momentum) that does not always correspond to a singlet whereas the AIM is always assumed to be spin averaged (see Eq. (\ref{singlet})). We therefore introduce a second step that accounts for the spin averaging. It corresponds to the energy difference between on the one hand the isolated atom with interpolated charge and averaged spin $E_{A}^{0,\overline{S}}(Q_{A})$ and on the other hand $E_{A}^{0}(Q_{A})$,
\small
\begin{eqnarray}
\Delta E_{A}^{\overline{S}} = E_{A}^{0,\overline{S}}(Q_{A}) - E_{A}^{0}(Q_{A}).
\label{spinaveragingstep}
\end{eqnarray}
\normalsize
$E_{A}^{0,\overline{S}}(Q_{A})$  is obtained as the interpolated value between the energies of the spin-averaged isolated atoms with an integer number of electrons,
\begin{eqnarray}
E_{A}^{0,\overline{S}}(Q_{A}) = a E_{A}^{0,\overline{S}}(N+1) + (1-a) E_{A}^{0,\overline{S}}(N),
\end{eqnarray}
and the $E_{A}^{0,\overline{S}}(N)$ are calculated from the following formula:
\begin{eqnarray}
E_{A}^{0,\overline{S}}(N)&=& \sum_{A}\sum_{ij} (t+V_{A}^{ext})_{ij} (\rho_{A}^{0})_{ij}  + \frac{1}{2}\sum_{ijkl} V_{ijkl}\left( (\rho_{A}^{0})_{ik}(\rho_{A}^{0})_{jl} - \frac{1}{2} (\rho_{A}^{0})_{il}(\rho_{A}^{0})_{jk} \right), 
\end{eqnarray}
where $\rho_{A}$ is the spin-summed 1DM of the isolated atom with N electrons.  

In the third step, charge reorganization takes place which corresponds to the difference between the self-
energy of the of the atomic 1DM fragment and the energy of the isolated atom with interpolated charge and averaged spin,
\begin{eqnarray}
\Delta E_{A}^{CR} = E^{self}_{A} - E_{A}^{0,\overline{S}}(Q_{A}).
\end{eqnarray}
Finally, the promotion energy is retrieved as the sum of three terms as shown below:
\begin{eqnarray}
E_{A}^{prom} = \Delta E_{A}^{CT} + \Delta E_{A}^{\overline{S}} + \Delta E_{A}^{CR}.
\label{bornhaber}
\end{eqnarray}
Although one may point out that there is some arbitrariness in the order chosen for the processes, the decomposition in Eq. (\ref{bornhaber}) might give a general idea about their relative importance.

\subsection{Bond energies}

Since the promotion and interaction energies depend on the AIM definition, they are not directly observable. The sum of all promotion and interaction energies, i.e. the atomization energy $\Delta E^{at}$ of the molecule, can be compared to experiment:
\begin{eqnarray}
\Delta E_{}^{at}= E^{mol}-\sum_{A}E_{A}^{0}= \sum_{A} E^{prom}_{A}+\sum_{A<B}E^{int}_{AB}.
\end{eqnarray}
In order to get a quantity that is more in line with the chemical concept of a ``bond energy'', it would be convenient to interpret the atomization of a molecule strictly in terms of atom pairs. Therefore, the atomic promotion energies ($E_{A}^{prom}$) are attributed to the atom pairs ($A < B$), 
\begin{eqnarray}
E_{AB}^{prom}= E_{A}^{prom}\left( \frac{E_{AB}^{int}}{\sum_C E_{AC}^{int}} \right) + E_{B}^{prom}\left( \frac{E_{AB}^{int}}{\sum_C E_{BC}^{int}} \right),
\end{eqnarray}
and Hartree-Fock bond energies are derived,
\begin{eqnarray}
E_{AB}^{bond}=  E_{AB}^{int}+E_{AB}^{prom},
\label{bondenergieshere}
\end{eqnarray}
that reproduce exactly the Hartree-Fock atomization energy
\begin{eqnarray}
\Delta E_{(HF)}^{at}=\sum_{A < B} E_{AB}^{bond}. 
\end{eqnarray}
These bond energies can be compared to average bond dissociation energies.

\section{Consistent decomposition of molecular properties \label{secdifferent}}

In the introduction we already noted that the current approach to calculate self-energies and interaction energies for 1DM fragments is unambiguous. That is not the case for the widely adopted approach to partition the molecular energy using numerical integration of the molecular integrals weighted by the atomic weight functions. In this section we explore the exemplary cases of two components of the atomic self-energy (see Eq. (\ref{selfenergy})): the atom-condensed kinetic energy and the atom-condensed Fock energy.

\subsection{Atom-condensed kinetic energy}

The strategy adopted in the current paper is to calculate energy components of the molecular fragments, rather than to fragment the molecular energy. For the kinetic energy in particular, this can have important consequences. The kinetic energy t$_{A}$ of the single-index atomic density matrix is calculated in Eqs. (\ref{selfenergy}- \ref{definitions}) in a finite basis set, but the corresponding expression in $\bm{r}$-space is: 
\begin{eqnarray}
t_{A} = -\frac{1}{2} \int d\bm{r} \left.  \nabla^2 \rho_{A}(\bm{r},\bm{r'}) \right|_{\bm{r}=\bm{r'}},
\label{eqtfrag}
\end{eqnarray}
where the notation $|_{\bm{r}=\bm{r'}}$ indicates that $\bm{r'}$ is replaced by $\bm{r}$ after the action of $\nabla^2_{(\bm{r})}$ on $\rho_{A}(\bm{r},\bm{r'})$ but before the integration is carried out. Note that different but equivalent representations of the kinetic energy operator exist,  \textit{i.e.} 
\begin{eqnarray}
t_{A}=\frac{1}{2} \int d\bm{r} \left. \nabla \cdot \nabla' \rho_{A}(\bm{r},\bm{r'}) \right|_{\bm{r}=\bm{r'}},
\label{twokin}
\end{eqnarray}
should give the same result as in Eq. (\ref{eqtfrag}). This is indeed fulfilled for the present formulation in terms of atomic density matrices, as follows directly from partial integration by generalizing a well known relationship for the kinetic energy of the molecular 1DM,
\begin{eqnarray}
 \frac{1}{2} \int d\bm{r}\left. \nabla \cdot \nabla' \rho_{}(\bm{r},\bm{r'}) \right|_{\bm{r}=\bm{r'}}=-\frac{1}{2} \int d\bm{r} \left.  \nabla^2 \rho_{}(\bm{r},\bm{r'}) \right|_{\bm{r}=\bm{r'}},
\end{eqnarray}
to its fragments. In contrast, the expression for the fragmentation of the molecular kinetic energy with Hirshfeld-I weight functions clearly depends on the representation of the kinetic energy operator \cite{ayers2002,anderson2010}, since in general
\begin{eqnarray}
t_{A}^{h} = -\frac{1}{2} \int  d\bm{r} \left. w_{A}(\bm{r})  \nabla^2 \rho(\bm{r},\bm{r'}) \right|_{\bm{r}=\bm{r'}} 
\label{eqtenfrag}
\end{eqnarray}
yields a result that differs from 
\begin{eqnarray}
t_{A}^{h'} = \frac{1}{2} \int  d\bm{r} \left. w_{A}(\bm{r})  \nabla \cdot \nabla' \rho(\bm{r},\bm{r'}) \right|_{\bm{r}=\bm{r'}}.
\label{eqtenfrag2}
\end{eqnarray}
Only in special cases, e.g. when QTAIM weight functions are used, do Eq. (\ref{eqtenfrag}) and Eq. (\ref{eqtenfrag2}) coincide. This shows that a naive fragmentation of the molecular energy by introducing weight functions in the molecular integrals is ambiguous by nature, in contrast to the fragmentation of the molecular 1DM and the calculation of the associated energy components.

It is shown in the appendix that the $\bm{r}$-space expressions for $t_{A}$ (Eqs.~(\ref{eqtfrag}) and (\ref{twokin}) ) and $t_{A}^{h}$ (Eq.~(\ref{eqtenfrag}) ) are mathematically identical. Since Eq. (\ref{eqtenfrag}) does not contain the action of the kinetic energy operator on an atomic weight function it can be calculated quite easily using 3D numerical integration on a spatial grid. This provides an indication of the error induced by using the finite basis set expression in Eq. (\ref{selfenergy}) instead of the full $\bm{r}$-space expression in Eq. (\ref{eqtfrag}).

\subsection{Atom-condensed Fock energy}

The Fock operator provides another example of a nonlocal operator that causes problems in the common approach, where atomic weight functions are inserted in the molecular Fock integrals. Indeed, there are several possibilities for the fragmentation of the Fock energy,
\begin{eqnarray}
E^{Fock}=-\frac{1}{2} \int d\bm{r} d\bm{r'} \phantom{.} \frac{[\rho(\bm{r},\bm{r'})]^{2}}{|\bm{r}-\bm{r'}|}.
\end{eqnarray}
Depending on the place where the decomposition of unity is inserted, one could have 
\begin{eqnarray}
E^{Fock} &=& \sum_{AB}-\frac{1}{2} \int d\bm{r}d\bm{r'} \phantom{.} w_A(\bm{r})w_B(\bm{r'}) \phantom{.} \frac{[\rho(\bm{r},\bm{r'})]^{2}}{|\bm{r}-\bm{r'}|}  \nonumber \\
 &=& \sum_{AB}-\frac{1}{4} \int d\bm{r}d\bm{r'} \phantom{.} [w_A(\bm{r})w_B(\bm{r})  + w_A(\bm{r'})w_B(\bm{r'})] \frac{[\rho(\bm{r},\bm{r'})]^{2}}{|\bm{r}-\bm{r'}|}   \\ 
& \vdots & \nonumber 
\end{eqnarray}
Only one fragmentation is consistent with the underlying partitioning in Eq. (\ref{rhoA}) of the molecular 1DM into single-index atomic density matrices, 
\begin{eqnarray}
E^{Fock} &=& \sum_{AB}-\frac{1}{2} \int  \frac{\rho_{A}(\bm{r},\bm{r'})\rho_{B}(\bm{r},\bm{r'})}{|\bm{r}-\bm{r'}|}  \nonumber \\
&=&\sum_{AB}-\frac{1}{8} \int d\bm{r}d\bm{r'}  \frac{}{} \phantom{.}[w_A(\bm{r})w_B(\bm{r}) + w_A(\bm{r'})w_B(\bm{r'})  \nonumber \\ && +  \frac{}{} w_A(\bm{r})w_B(\bm{r'})  + w_B(\bm{r})w_A(\bm{r'})] \phantom{.} \frac{[\rho(\bm{r},\bm{r'})]^{2}}{|\bm{r}-\bm{r'}|} . 
\label{fockinter}
\end{eqnarray}
Note that the Fock contributions with indices (A $=$ B) are attributed to $E^{self}_{A}$, while the Fock contributions with indices (A $\neq$ B) are attributed to $E^{int}_{AB}$.

\section{A fast decomposition in Hilbert space}

In a previous paper \cite{vanfleteren2010}, it was observed that the molecular 1DM fragments show a satisfactory basis set convergence when they are expressed in one-electron Hilbert space. To avoid cumbersome numerical integrations in $\bm{r}$-space, atomic self-energies and interaction energies are calculated efficiently in one-electron Hilbert space (see Eqs. (\ref{selfenergy}) and (\ref{subinter})). The widespread alternative strategy to fragment the molecular energy (by inserting atomic weight functions in the molecular expressions) requires that the atom-condensed energy integrals are computed numerically on a large spatial grid. In this section we explore the computational consequences for two components of the atomic self-energy (see Eq. (\ref{selfenergy})): the atom-condensed kinetic energy and the atom-condensed Fock energy.

\subsection{Atom-condensed Kinetic energy}

When the kinetic energy is calculated by numerical integration of the molecular energy integrals weighted by the atomic weight functions (see Eq. \ref{eqtenfrag}),
\begin{eqnarray}
t_A^{h} &=&-\frac{1}{2} \sum_{i}d_i \int d{\bm{r}} w_A(\bm{r}) \left[ \phi_i(\bm{r}) \left( \nabla^{2} \phi_i(\bm{r}) \right) \frac{}{}\right],
\end{eqnarray}
it is clear that only few integrals have to be computed since nonzero contributions are obtained only for occupied MO's. This is an important computational advantage over using an atomic density matrix as in Eq. (\ref{selfenergy}-\ref{definitions}), to which all orbitals in the basis contribute. However, experience shows that the numerical computation of the integrals above requires quite large integration grids whereas the atomic overlap integrals can be computed with sufficient precision using more modest grids. As will be shown, there is a trade-off between either computing numerically fewer integrals that require larger grids and more atomic overlap integrals that are individually computed much easier.

The strategy adopted in the current paper is to calculate energy components via matrix manipulations in one-electron Hilbert space (See Eqs. (\ref{selfenergy}) and (\ref{subinter}) ). Using Eq. (\ref{dens1}), the precise expression for the kinetic energy is, 
\begin{eqnarray}
t_A &=& \frac{1}{2}\sum_{ij} (d_i +d_j) t_{ij} S^{A}_{ij}. 
\end{eqnarray}
A non-zero contribution is obtained if either $d_i$ or $d_j$ is nonzero. This implies that the summation essentially runs over all i and j and that all of the following integrals must be computed:
\begin{eqnarray}
S^{A}_{ij}&=& \int d \bm{r} \phi_i (\bm{r}) w_A (\bm{r})\phi_j(\bm{r}) \nonumber \\
t_{ij}&=& -\frac{1}{2}\int  \phi_{i}^{}(\bm{r}) \left(\nabla^{2} \phi_{j}(\bm{r})\right) d\bm{r}  \nonumber 
\end{eqnarray}
However, the $t_{ij}$ can be calculated analytically and moderate grids suffice to construct the $S^{A}_{ij}$. 

\subsection{Atom-condensed Fock energy}

For the partitioning of e.g. the Fock-energy, the considerations of the previous subsection are slightly more outspoken than for the kinetic energy. Using the strategy of numerical integration,
\begin{eqnarray}
E^{Fock}_{AB} &=&-\frac{1}{8} \sum_{ij} d_i d_j \int d\bm{r}d\bm{r'}  \frac{}{} \phantom{.}[w_A(\bm{r})w_B(\bm{r}) + w_A(\bm{r'})w_B(\bm{r'})  \nonumber \\ && +  \frac{}{} w_A(\bm{r})w_B(\bm{r'})  + w_B(\bm{r})w_A(\bm{r'})] \phantom{.} \frac{\phi_{i}^{}(\bm{r})\phi_{i}^{}(\bm{r'})\phi_{j}^{}(\bm{r})\phi_{j}^{}(\bm{r'}) }{|\bm{r}-\bm{r'}|} , 
\end{eqnarray}
the number of integrals that has to be computed is limited to the square of the number of occupied orbitals. On the other hand, the cost of numerical integration is squared with respect to the situation for the kinetic energy since the integrals run over $\bm{r}$ and $\bm{r'}$.

When the Fock energy is partitioned via matrix manipulations in one-electron Hilbert space, the precise expression is: 
\begin{eqnarray}
E^{Fock}_{AB} &=& -\frac{1}{16} \sum_{ijkl} (d_i+d_l)(d_j+d_k) V_{ijkl} S^{A}_{il}S^{B}_{jk}, 
\end{eqnarray}
where $V_{ijkl}$ is calculated analytically. The summation runs over all ijkl, but the $S^{A}_{il}$ are calculated on a moderate grid that runs over only $\bm{r}$. Except for large systems, the approach presented in this paper is more appealing from a computational point of view. Note that $E^{Fock}_{AB}$ is part of $E^{self}_{A}$ when ($A = B$), while it is part of $E^{int}_{AB}$ when ($A \neq B$).

\section{Computational methods}

The formalism described in the sections \ref{single-atom}-\ref{energy decomposition} was applied to the set of 25 molecules listed in table \ref{testset}, representing a variety of chemical bonds. All molecules have a singlet ground state, apart from O$_{2}$, for which we consider the singlet (excited) state.

The molecular 1DM was calculated at the restricted open-shell Hartree-Fock (ROHF) \cite{roothaan1951,pople1954,mcweeny1968} level of theory  using the (cartesian) Aug-cc-pVTZ basis, including a geometry optimization. 
All molecular one-electron density matrices were calculated with the Gaussian 03 program \cite{g03}.

\begin{table} [ht!]
\centering
\begin{tabular}{|l|l|l|l|l|l|l|} 
\hline 
H$_2$	&	N$_2$	&	HCl	&	CH$_3$OH	&	H$_2$O	\\
F$_2$	&	LiH	&	CH$_4$	&	H$_2$CO	&	(H$_2$O)$_2$	\\
Cl$_2$	&	LiF	&	CH$_3$CH$_3$	&	CHOOH	&	H$_2$O$_2$	\\
Li$_2$	&	NaCl	&	CH$_2$CH$_2$	&	CO$_2$	&	NH$_3$	\\
O$_2$	&	HF	&	CHCH	&	CO	&	N$_2$H$_4$	\\
\hline 
\end{tabular} 
\caption{List of molecules in the test set.}
\label{testset}
\end{table}

The partitioning scheme was implemented using the atomic weight functions $h_{A}(\bm{r})$ in Eq.(\ref{eq5}) from a Hirshfeld-I analysis. 
To be consistent with our previous work on the Hirshfeld-I compatible 1DM partitioning \cite{vanfleteren2010},  the iterative Hirshfeld weights $h_{A}(\bm{r})$
and the $S^A_{ij}$ 
coefficients derived from these weights were calculated on atom-centered grids, using a logarithmic radial grid of 100 points with $r_{\text{min}}=10^{-6}$ \r{A}  and $r_\text{max}=20 $ \r{A} , and the 170-point Lebedev angular grids, \cite{lebedev1975,lebedev1976,lebedev1977,lebedev1992,lebedev1995,lebedev1999} with each radial shell given a randomized orientation. The sum rule of Eq.(\ref{sumrule}) was fulfilled with a reasonable precision of $10^{-3}$ except for some very diffuse virtual orbitals (as we work in an augmented basis set), where the sum rule is fulfilled with less precision ($10^{-2}$) because the grid is too localized. However, as these orbitals are of little relevance in the calculation of the energies, this does not significantly affect the outcome of our analysis. In order to reproduce the molecular energy in Eq. (\ref{reconstruction}) exactly, the sum rule of Eq.(\ref{sumrule}) was enforced by a renormalisation of the atom-condensed overlap matrices.

\section{Results}

\subsection{Energy decomposition using the single-atom density matrices}

Table {\ref{totalpromotion}} displays the total promotion and interaction energies belonging to the (single-index) density matrix fragments (see section \ref{sec-en-decom}). 
\linespread{1.2}
\begin{table} [H]
\centering
\footnotesize
\begin{tabular}{|r|r|r|r|r|r|r|r|r|r|r|r|}
\linespread{1.2}
   & 	  E$^{prom}_{(HF)}$ & E$^{int}_{(HF)}$ & -$\Delta E_{(HF)}^{at}$ & -$\Delta E_{CC}^{at}$   & &     E$^{prom}_{(HF)}$ & E$^{int}_{(HF)}$ & -$\Delta E_{(HF)}^{at}$ & -$\Delta E_{CC}^{at}$\\
\hline 
H$_2$	&		0.18	&	-0.31	&	-0.13	&	-0.17	&	CH$_2$-CH$_2$	&		2.24	&	-2.93	&	-0.69	&	-0.88	\\
F$_2$		&	0.75	&	-0.71	&	0.04	&	-0.05	&	CHCH		&	1.84	&	-2.32	&	-0.48	&	-0.63	\\
Cl$_2$		&	0.56	&	-0.60	&	-0.04	&	-0.09	&	CH$_3$OH	&		2.25	&	-2.86	&	-0.60	&	-0.80	\\
Li$_2$		&	0.11	&	-0.12	&	-0.01	&	-0.04	&	H$_2$CO		&	1.85	&	-2.28	&	-0.42	&	-0.58	\\
O$_2$		&	1.37	&	-1.35	&	0.01	&	-0.13	&	CHOOH		&	2.94	&	-3.48	&	-0.55	&	-0.78	\\
N$_2$		&	1.60	&	-1.79	&	-0.19	&	-0.34	&	CO$_2$		&	2.61	&	-3.02	&	-0.41	&	-0.60	\\
LiH		&	0.35	&	-0.40	&	-0.05	&	-0.09	&	CO		&	1.29	&	-1.58	&	-0.29	&	-0.40	\\
LiF		&	0.31	&	-0.46	&	-0.15	&	-0.22	&	H$_2$O		&	1.06	&	-1.32	&	-0.26	&	-0.36	\\
NaCl		&	0.05	&	-0.16	&	-0.12	&	-0.15	&	(H$_2$O)$_2$	&	2.16	&	-2.68	&	-0.52	&	-0.73	\\
HF		&	0.45	&	-0.61	&	-0.16	&	-0.22	&	H$_2$O$_2$	&		1.79	&	-2.02	&	-0.23	&	-0.41	\\
HCl		&	0.35	&	-0.48	&	-0.13	&	-0.17	&	NH$_3$		&	1.48	&	-1.80	&	-0.32	&	-0.46	\\
CH$_4$		&	1.39	&	-1.91	&	-0.53	&	-0.66	&	N$_2$H$_4$	&		2.72	&	-3.16	&	-0.43	&	-0.68	\\
CH$_3$CH$_3$		&	2.61	&	-3.50	&	-0.89	&	-1.12	&		&		&			&		&		\\
\hline
\end{tabular} 
\linespread{1.5}
\caption{Total promotion and interaction energies (E$^{prom}_{(HF)}$ and E$^{int}_{(HF)}$) calculated at the ROHF/Aug-cc-pVTZ level of theory. For comparison with the HF atomization energies $\Delta E_{(HF)}^{at}$, the CCSD(T) atomization energies $\Delta E_{CC}^{at}$ are also presented. All energies are in Hartree.}
\label{totalpromotion}
\end{table}
\linespread{1.5}
The sum of the total promotion and interaction energies yields (minus) the Hartree-Fock atomization energy -$\Delta E_{(HF)}^{at}$; in most cases this represents about 3/4 of the atomization energy calculated at the CCSD(T) \cite{bartlett1978, pople1978, cizek1969, purvis1982, scuseria1988, scuseria1989} level of theory  -$\Delta E_{CC}^{at}$. Note that in some cases the Hartree-Fock atomization energy is far from satisfying, e.g. it predicts that $F_{2}$ has a higher energy then two isolated F-atoms in their ground state. The magnitude of the promotion and interaction energies is within the range of typical IQA (Interacting Quantum Atoms) \cite{blanco2005,pendas2006a,pendas2006b} values, studied by Pendas et al. \cite{pendas2007} for different methods including the Hirshfeld AIM method.

\begin{table} [H]
\linespread{1.2}
\centering
\footnotesize
\begin{tabular}{|r|r|r|r|r|r|r|r|r|r|r|r|r|r|r|r|}
\hline
    & at & E$_{A}^{0}$ \hspace{2mm} & Q$_{A}$ \hspace{3mm} & $\Delta E_{A}^{CT}$ & $\Delta E_A^{\overline{S}}$  &  $\Delta E_{A}^{CR}$ & E$_{A}^{prom}$ & & at & E$_{A}^{0}$ \hspace{2mm} & Q$_{A}$ \hspace{3mm} & $\Delta E_{A}^{CT}$  & $\Delta E_A^{\overline{S}}$ &  $\Delta E_{A}^{CR}$ & E$_{A}^{prom}$\\
\hline 
H$_2$	&	H	&	-0.50	&	0.00	&	0.00	&	0.16	&	-0.07	&	0.09	&	H$_2$CO	&	C	&	-37.69	&	0.47	&	0.18	&	0.26	&	0.63	&	1.07	\\
F$_2$	&	F	&	-99.40	&	0.00	&	0.00	&	0.24	&	0.14	&	0.38	&		&	H	&	-0.50	&	0.01	&	0.01	&	0.15	&	-0.02	&	0.13	\\
Cl$_2$	&	Cl	&	-459.48	&	0.00	&	0.00	&	0.15	&	0.13	&	0.28	&		&	O	&	-74.81	&	-0.49	&	0.01	&	0.33	&	0.18	&	0.52	\\
Li$_2$	&	Li	&	-7.43	&	0.00	&	0.00	&	0.06	&	0.00	&	0.06	&	CHOOH	&	C	&	-37.69	&	0.95	&	0.38	&	0.18	&	0.90	&	1.45	\\
O$_2$	&	O	&	-74.81	&	0.00	&	0.00	&	0.45	&	0.23	&	0.68	&		&	O	&	-74.81	&	-0.78	&	0.01	&	0.26	&	0.32	&	0.59	\\
N$_2$	&	N	&	-54.40	&	0.00	&	0.00	&	0.59	&	0.21	&	0.80	&		&	O	&	-74.81	&	-0.67	&	0.01	&	0.29	&	0.19	&	0.49	\\
LiH	&	Li	&	-7.43	&	0.97	&	0.19	&	0.00	&	0.13	&	0.32	&		&	H	&	-0.50	&	0.00	&	0.00	&	0.16	&	-0.03	&	0.13	\\
	&	H	&	-0.50	&	-0.97	&	0.01	&	0.01	&	0.00	&	0.02	&		&	H	&	-0.50	&	0.51	&	0.25	&	0.08	&	-0.06	&	0.27	\\
LiF	&	Li	&	-7.43	&	0.98	&	0.19	&	0.00	&	0.18	&	0.37	&	CO$_2$	&	C	&	-37.69	&	1.23	&	0.60	&	0.14	&	0.99	&	1.73	\\
	&	F	&	-99.40	&	-0.98	&	-0.05	&	0.00	&	-0.01	&	-0.06	&		&	O	&	-74.81	&	-0.62	&	0.01	&	0.30	&	0.12	&	0.44	\\
NaCl	&	Na	&	-161.86	&	0.96	&	0.18	&	0.00	&	0.11	&	0.28	&	CO	&	C	&	-37.69	&	0.27	&	0.11	&	0.28	&	0.31	&	0.69	\\
	&	Cl	&	-459.48	&	-0.96	&	-0.09	&	0.01	&	-0.15	&	-0.23	&		&	O	&	-74.81	&	-0.27	&	0.01	&	0.38	&	0.21	&	0.60	\\
HF	&	H	&	-0.50	&	0.55	&	0.27	&	0.08	&	-0.08	&	0.28	&	H$_2$O	&	O	&	-74.81	&	-0.95	&	0.02	&	0.21	&	0.28	&	0.51	\\
	&	F	&	-99.40	&	-0.55	&	-0.03	&	0.11	&	0.09	&	0.17	&		&	H	&	-0.50	&	0.48	&	0.24	&	0.08	&	-0.04	&	0.27	\\
HCl	&	H	&	-0.50	&	0.26	&	0.13	&	0.12	&	-0.05	&	0.20	&	(H$_2$O)$_2$	&	O	&	-74.81	&	-0.98	&	0.02	&	0.21	&	0.28	&	0.51	\\
	&	Cl	&	-459.48	&	-0.26	&	-0.02	&	0.10	&	0.07	&	0.15	&		&	H	&	-0.50	&	0.49	&	0.24	&	0.08	&	-0.04	&	0.29	\\
CH$_4$	&	C	&	-37.69	&	-0.47	&	-0.01	&	0.36	&	0.38	&	0.73	&		&	O	&	-74.81	&	-0.97	&	0.02	&	0.21	&	0.30	&	0.53	\\
	&	H	&	-0.50	&	0.12	&	0.06	&	0.14	&	-0.03	&	0.16	&		&	H	&	-0.50	&	0.47	&	0.24	&	0.08	&	-0.04	&	0.27	\\
CH$_3$CH$_3$	&	C	&	-37.69	&	-0.26	&	-0.01	&	0.34	&	0.48	&	0.82	&		&	H	&	-0.50	&	0.50	&	0.25	&	0.08	&	-0.05	&	0.28	\\
	&	H	&	-0.50	&	0.09	&	0.04	&	0.15	&	-0.03	&	0.16	&	H$_2$O$_2$	&	O	&	-74.81	&	-0.43	&	0.01	&	0.34	&	0.28	&	0.63	\\
	&	H	&	-0.50	&	0.09	&	0.04	&	0.15	&	-0.03	&	0.16	&		&	H	&	-0.50	&	0.43	&	0.22	&	0.08	&	-0.03	&	0.26	\\
CH$_2$CH$_2$	&	C	&	-37.69	&	-0.21	&	0.00	&	0.33	&	0.48	&	0.80	&	NH$_3$	&	N	&	-54.40	&	-1.02	&	0.08	&	0.43	&	0.27	&	0.78	\\
	&	H	&	-0.50	&	0.11	&	0.05	&	0.14	&	-0.03	&	0.16	&		&	H	&	-0.50	&	0.34	&	0.17	&	0.10	&	-0.04	&	0.23	\\
CHCH	&	C	&	-37.69	&	-0.21	&	0.00	&	0.33	&	0.43	&	0.76	&	N$_2$H$_4$	&	N	&	-54.40	&	-0.58	&	0.04	&	0.50	&	0.37	&	0.91	\\
	&	H	&	-0.50	&	0.21	&	0.11	&	0.12	&	-0.06	&	0.16	&		&	H	&	-0.50	&	0.28	&	0.14	&	0.11	&	-0.03	&	0.22	\\
CH$_3$OH	&	C	&	-37.69	&	0.10	&	0.04	&	0.30	&	0.59	&	0.93	&		&	H	&	-0.50	&	0.30	&	0.15	&	0.11	&	-0.03	&	0.23	\\
	&	H	&	-0.50	&	0.07	&	0.03	&	0.15	&	-0.03	&	0.15	&		&		&		&		&		&		&		&		\\
	&	H	&	-0.50	&	0.04	&	0.02	&	0.15	&	-0.02	&	0.15	&		&		&		&		&		&		&		&		\\
	&	O	&	-74.81	&	-0.68	&	0.01	&	0.29	&	0.30	&	0.60	&		&		&		&		&		&		&		&		\\
	&	H	&	-0.50	&	0.44	&	0.22	&	0.09	&	-0.04	&	0.27	&		&		&		&		&		&		&		&		\\
\hline
\end{tabular} 
\linespread{1.5}
\caption{Charge transfer, spin-averaging and charge redistribution energies ($\Delta E_{A}^{CT}$, $\Delta E_A^{\overline{S}}$ and $\Delta E_{A}^{CR}$, in Hartree) for the molecular fragments of some small molecules calculated at the ROHF/Aug-cc-pVTZ level of theory. These are the components of the atomic promotion energies E$_{A}^{prom}$. The reference energy, E$_{A}^{0}$, is the energy of the neutral isolated atom calculated at the ROHF/Aug-cc-pVTZ level of theory. $Q_{A}$ is the charge of the single-index atom.}
\label{atomicpromotion}
\linespread{1.5}
\end{table}
\normalsize
\linespread{1.5}

Table {\ref{atomicpromotion}} displays the charge transfer (CT), spin averaging (S) and the charge redistribution (CR) components of the individual atomic promotion energies. The current scheme is based on a partitioning of the 1DM. To calculate the energy components of the 1DM fragments, it is necessary to specify the electron spin of the fragments. For singlet molecules, the current scheme averages the electron spin of the 1DM fragments. In an attempt to avoid the spin-averaging step and get lower atomic promotion energies, one could implement an alternative 1DM partitioning scheme and request that e.g. the hydrogen atoms keep their doublet structure in a H$_2$ molecule. However, such requirement would lead to delocalized electron densities for the AIMs and would exclude consistency with the (well-localized) Hirshfeld-I model. Since the localization of the AIM densities is a necessary condition to produce chemically meaningful results within the IQA approach \cite{pendas2007}, it is clear that the spin must be averaged at least to some degree. Note that the problem related to the spin is not addressed by the common energy decomposition schemes that are based on a partitioning of the molecular expectation values using AIM weight functions. In these schemes, the underlying electronic structure is not explicitly treated (and possibly not consistent).
 Energetically, it appears that the spin averaging energy defined in Eq. (\ref{spinaveragingstep}) is of similar importance as the charge redistribution energy. E.g. the average of the absolute values in Table {\ref{atomicpromotion}} for $\Delta E^{CT}_{A}$, $\Delta E^{\overline{S}}_{A}$ and $\Delta E^{CR}_{A}$ is 0.09, 0.19 and 0.19 Hartree respectively. The charge redistribution (CR) energies are mostly positive, although small negative values are found for the hydrogen atoms. At first sight, one would expect these CR-energies to be positive for variational reasons, since it is based on the atomic HF-Hamiltonian, for which the isolated atom 1DM with fractional charge should be optimal. However, $\rho_{A}$ is not necessarily N-representable \cite{verstichel2009}, so it is variationally not sufficiently constrained. It is nevertheless remarkable that these negative values are rather small. 

In Table {\ref{bondingenergies}} the interaction energies between the molecular fragments are presented. The Fock component is listed separately,  
\begin{eqnarray}
F_{AB}^{int}=\sum_{ijkl} V_{ijkl}\left( \frac{1}{2} (\rho_{A})_{il} (\rho_{B})_{jk} \right) \label{ECM1}, 
\end{eqnarray}
as it seems to be a robust indicator of the covalent character of that interaction. It clearly singles out the ionic species with a small value, whereas for the (covalent) homonuclear diatomic molecules in the analysis, it correlates linearly with the interaction energy (R$^{2}$=0.997). In contrast to the SEDI (bond order) index \cite{wiberg1968, bader1975, giambiagi1975, mayer1983, fulton1993, angyan1994, mayer2004, fradera1999, mayer2007, ponec2005} the value is also low for the covalent but weakly bound Li$_{2}$. 

The bond energies are also listed in Table {\ref{bondingenergies}}. They do depend on the choice of reference atoms, but chemists are more familiar with their typical magnitudes, and much chemical reasoning is based on this type of quantity. The linear correlation between the interaction energies and the bond energies is not strong (R$^{2}$=0.6). Both quantities can single out chemically not bonded atoms (having very small values) and for both quantities similar values are obtained for chemically similar bonds. The C-H bond has interaction energies between 0.45 Hartree (in CH$_{4}$) and 0.40 Hartree in (CHOOH) and bond energies between 0.12 and 0.8 Hartree. Interaction energies for the single, double and triple C-C bonds in ethane, ethene and ethyn are respectively -0.58, -0.98 and -1.38 Hartree, while the corresponding bond energies are 0.10, 0.17 and 0.25 Hartree.  The interaction energies are theoretically appealing in the sense that they follow immediately from the density matrix partitioning. They should not be confused with the bond energies, which are more in line with chemical intuition but require the introduction of isolated reference atoms. The analysis presented in this work is particularly useful to investigate substituent effects in molecules. When an atom or functional group is substituted in the molecule, the strength of all bonds in the molecule is affected. The bond energies defined in Eq. (\ref{bondenergieshere}) are a measure for this effect. E.g. it is clear from the entries in Table {\ref{bondingenergies}} that the C-H bond in ethene (CH$_{2}$CH$_{2}$) is stronger than the C-H bond in formaldehyde (CH$_2$O) and formic acid (CHOOH). It is also possible to compare the interactions between atoms that have no chemical bond in the molecule. Although we recognize that it is possible to get similar information from other schemes that were developed to partition the Hartree-Fock energy, we stress the consistency of this approach, where different energy terms are derived from the same, symmetrical, atom condensed density matrices. Without this mathematical consistency, the results are necessarily ambiguous.

\linespread{1.2}

\begin{table}[H]
\footnotesize 
\centering

\begin{tabular}{|l|l|r|r|r|r|r|l|l|r|r|r|r|r|}
\hline
    & bond & SEDI$_{AB}$ \hspace{0mm} & F$^{int}_{AB}$ \hspace{2mm} &  E$^{int}_{AB}$ \hspace{2mm} & E$^{prom}_{AB}$ \hspace{0mm} & E$^{bond}_{AB}$ \hspace{1mm} & & bond & b.o.$_{AB}$ \hspace{0mm} & F$^{int}_{AB}$ \hspace{2mm} &  E$^{int}_{AB}$ \hspace{2mm} & E$^{prom}_{AB}$ \hspace{0mm} & E$^{bond}_{AB}$ \hspace{1mm} \\
\hline 
H$_2$	&	H-H	&	1.00	&	-0.26	&	-0.31	&	-0.18	&	-0.13	&	H$_2$CO	&	C-O	&	2.29	&	-0.77	&	-1.34	&	-1.15	&	-0.20	\\
F$_2$	&	F-F	&	1.69	&	-0.48	&	-0.71	&	-0.75	&	0.04	&		&	C-H	&	0.94	&	-0.29	&	-0.41	&	-0.32	&	-0.09	\\
Cl$_2$	&	Cl-Cl	&	1.84	&	-0.40	&	-0.60	&	-0.56	&	-0.04	&		&	O$\cdots$H	&	0.20	&	-0.03	&	-0.04	&	-0.03	&	-0.02	\\
Li$_2$	&	Li-Li	&	1.03	&	-0.10	&	-0.12	&	-0.11	&	-0.01	&	CHOOH	&	C-O	&	1.35	&	-0.47	&	-1.03	&	-0.92	&	-0.11	\\
O$_2$	&	O-O	&	2.77	&	-0.85	&	-1.35	&	-1.37	&	0.01	&		&	C-O	&	1.98	&	-0.70	&	-1.40	&	-1.21	&	-0.19	\\
N$_2$	&	N-N	&	3.36	&	-1.09	&	-1.79	&	-1.79	&	-1.79	&		&	C-H	&	0.85	&	-0.27	&	-0.40	&	-0.32	&	-0.08	\\
LiH	&	Li-H	&	0.26	&	-0.06	&	-0.40	&	-0.35	&	-0.05	&		&	O-H	&	0.75	&	-0.26	&	-0.62	&	-0.51	&	-0.12	\\
LiF	&	Li-F	&	0.27	&	-0.08	&	-0.46	&	-0.31	&	-0.15	&		&	O$\cdots$HC	&	0.17	&	-0.04	&	-0.04	&	-0.03	&	-0.02	\\
NaCl	&	Na-Cl	&	0.31	&	-0.06	&	-0.16	&	-0.05	&	-0.12	&		&	O$\cdots$HC	&	0.20	&	-0.04	&	-0.04	&	-0.02	&	-0.02	\\
HF	&	H-F	&	0.79	&	-0.27	&	-0.61	&	-0.45	&	-0.16	&		&	O$\cdots$HO	&	0.04	&	-0.01	&	-0.09	&	-0.07	&	-0.02	\\
HCl	&	H-Cl	&	1.13	&	-0.30	&	-0.48	&	-0.35	&	-0.13	&	CO$_2$	&	C-O	&	2.08	&	-0.74	&	-1.52	&	-1.31	&	-0.21	\\
CH$_4$	&	C-H	&	0.99	&	-0.30	&	-0.45	&	-0.33	&	-0.12	&		&	O$\cdots$O	&	0.46	&	-0.07	&	0.02	&	0.01	&	0.01	\\
	&	H$\cdots$H	&	0.07	&	-0.01	&	-0.02	&	-0.01	&	-0.01	&	CO	&	C-O	&	2.88	&	-0.95	&	-1.58	&	-1.29	&	-0.29	\\
CH$_3$CH$_3$	&	C-C	&	1.23	&	-0.39	&	-0.58	&	-0.48	&	-0.10	&	H$_2$O	&	O-H	&	0.85	&	-0.29	&	-0.69	&	-0.56	&	-0.13	\\
	&	C-H	&	0.95	&	-0.30	&	-0.43	&	-0.32	&	-0.12	&		&	H$\cdots$H	&	0.03	&	-0.01	&	0.06	&	0.05	&	0.01	\\
	&	C$\cdots$H	&	0.10	&	-0.02	&	-0.03	&	-0.02	&	-0.01	&	(H$_2$O)$_2$	&	O-H	&	0.78	&	-0.27	&	-0.68	&	-0.55	&	-0.13	\\
	&	C$_1$H$\cdots$HC$_1$	&	0.07	&	-0.01	&	-0.02	&	-0.01	&	-0.01	&		&	O-H	&	0.86	&	-0.29	&	-0.69	&	-0.56	&	-0.14	\\
CH$_2$CH$_2$	&	C-C	&	2.06	&	-0.63	&	-0.98	&	-0.81	&	-0.17	&		&	O-H	&	0.82	&	-0.28	&	-0.69	&	-0.56	&	-0.13	\\
	&	C-H	&	0.96	&	-0.30	&	-0.44	&	-0.32	&	-0.12	&		&	O$\cdots$O	&	0.11	&	-0.02	&	0.15	&	0.11	&	0.04	\\
	&	C$\cdots$H	&	0.15	&	-0.03	&	-0.04	&	-0.03	&	-0.01	&		&	O$^{--}$H	&	0.07	&	-0.02	&	-0.14	&	-0.11	&	-0.03	\\
	&	C$_1$H$\cdots$HC$_1$	&	0.06	&	-0.01	&	-0.01	&	-0.01	&	-0.01	&		&	O$\cdots$H	&	0.00	&	0.00	&	-0.07	&	-0.06	&	-0.01	\\
CH-CH	&	C-C	&	2.98	&	-0.88	&	-1.38	&	-1.13	&	-0.25	&		&	O$\cdots$H	&	0.00	&	0.00	&	-0.07	&	-0.05	&	-0.01	\\
	&	C-H	&	0.96	&	-0.30	&	-0.43	&	-0.33	&	-0.10	&	H$_2$O$_2$	&	O-O	&	1.76	&	-0.53	&	-0.77	&	-0.69	&	-0.08	\\
	&	C$\cdots$H	&	0.15	&	-0.02	&	-0.04	&	-0.03	&	-0.01	&		&	O-H	&	0.85	&	-0.29	&	-0.56	&	-0.49	&	-0.07	\\
	&	H$\cdots$H	&	0.01	&	0.00	&	0.01	&	0.00	&	0.00	&		&	O..H	&	0.10	&	-0.02	&	-0.08	&	-0.07	&	-0.01	\\
CH$_3$OH	&	C-O	&	1.44	&	-0.48	&	-0.79	&	-0.67	&	-0.12	&	NH$_3$	&	N-H	&	0.95	&	-0.31	&	-0.62	&	-0.51	&	-0.11	\\
	&	C-H	&	0.94	&	-0.29	&	-0.42	&	-0.32	&	-0.10	&	N$_2$H$_4$	&	N-N	&	1.70	&	-0.54	&	-0.76	&	-0.70	&	-0.07	\\
	&	C-H	&	0.93	&	-0.29	&	-0.41	&	-0.32	&	-0.10	&		&	N-H	&	0.94	&	-0.31	&	-0.54	&	-0.46	&	-0.09	\\
	&	O-H	&	0.85	&	-0.29	&	-0.62	&	-0.51	&	-0.11	&		&	N-H	&	0.92	&	-0.31	&	-0.54	&	-0.46	&	-0.08	\\
	&	O$\cdots$H	&	0.14	&	-0.03	&	-0.04	&	-0.03	&	-0.01	&		&	N$\cdots$H	&	0.11	&	-0.02	&	-0.07	&	-0.06	&	-0.01	\\
	&	O$\cdots$H	&	0.13	&	-0.03	&	-0.05	&	-0.03	&	-0.01	&		&	N$\cdots$H	&	0.11	&	-0.02	&	-0.08	&	-0.07	&	-0.01	\\
\hline
\end{tabular} 
\linespread{1.1}
\caption{Interaction energies (E$_{AB}^{int}$, in Hartree) for the molecular fragments of some small molecules calculated at the ROHF/Aug-cc-pVTZ level of theory. In the column labeled "bond", the notations "-", "$^{--}$" and "$\cdots$" indicate respectively a chemical bond, a hydrogen bond and the interaction between a pair of non-bonded atoms. For comparison, the Hirshfeld-I SEDI-index is included as a bond order indicator and the Fock part of the interaction energy (F$_{AB}^{int}$) seems to be a robust indicator of covalent interaction energy. Atomic promotion energies are attributed to the bonds (E$_{AB}^{prom}$) to derive Hartree-Fock bond energies (E$_{AB}^{bond}$). }
\label{bondingenergies}
\end{table}
\linespread{1.5}

\subsection{Consistent decomposition of molecular properties}

Table \ref{kineticenergies} displays the kinetic energies t$_{A}$ from the single-index atomic density matrices (see Eq. (\ref{selfenergy})) calculated with the matrix elements of the finite Aug-cc-pVTZ basis set. They are compared to the atom-condensed kinetic energies t$_{A}^{h}$ [Eq. (\ref{eqtenfrag})] and t$_{A}^{h'}$ [Eq. (\ref{eqtenfrag2})], obtained by numerical computation in $\bm{r}$-space of the molecular kinetic energy integral weighted by the Hirshfeld-I atomic weight functions.
We confirmed that all entries in table \ref{kineticenergies} are converged with respect to grid size. Note that for the homonuclear diatomic case, all values coincide, since there is only one way to partition the molecular kinetic energy over equivalent atoms. 

As discussed in section \ref{secdifferent} the atom-condensed kinetic energies t$_{A}^{h}$ and t$_{A}^{h'}$ differ significantly (e.g. a difference of 0.3 Hartree for C in CO) pointing to the inherent ambiguity of combining atomic weight functions with the kinetic energy operator. These ambiguities are absent when the molecular 1DM itself is partitioned. For infinite basis sets, values for the kinetic energy of the 1DM fragments t$_{A}$ should equal values for t$_{A}^{h}$. It is clear from the table that for the present Aug-cc-pVTZ basis set significant differences still occur. For example, the difference $|t_{C}^{h}-t_{C}|$ is more than 0.06 Hartree in CO. We checked for CO that this difference vanishes in the Aug-cc-pVQZ basis set ($|t_{C}^{h}-t_{C}|$ = 0.002 Hartree).  

\linespread{1.2}
\begin{table} [H]
\centering
\footnotesize
\begin{tabular}{|r|r|r|r|r|r|r|r|r|r|r|r|r|r|}
\hline
   & A & t$_{A}^{0}$&  t$_{A}$-t$_{A}^{0}$ &  t$_{A}^{h}$-t$_{A}^{0}$ &  t$_{A}^{h'}$-t$_{A}^{0}$ &      & A & t$_{A}^{0}$ &  t$_{A}$-t$_{A}^{0}$ &  t$_{A}^{h}$-t$_{A}^{0}$ &  t$_{A}^{h'}$-t$_{A}^{0}$    \\
\hline 
H$_2$	&	H	&	0.50	&	0.07	&	0.07	&	0.07	&	H$_2$CO	&	C	&	37.68	&	0.00	&	-0.08	&	-0.13	\\
F$_2$	&	F	&	99.40	&	-0.02	&	-0.02	&	-0.02	&		&	H	&	0.50	&	0.12	&	0.12	&	0.25	\\
Cl$_2$	&	Cl	&	459.44	&	0.01	&	0.01	&	0.01	&		&	O	&	74.80	&	0.19	&	0.26	&	0.06	\\
Li$_2$	&	Li	&	7.43	&	0.00	&	0.00	&	0.00	&	CHOOH	&	C	&	37.68	&	-0.21	&	-0.33	&	-0.38	\\
O$_2$	&	O	&	74.80	&	-0.04	&	-0.04	&	-0.04	&		&	O	&	74.80	&	0.45	&	0.52	&	0.27	\\
N$_2$	&	N	&	54.39	&	0.10	&	0.10	&	0.10	&		&	O	&	74.80	&	0.31	&	0.37	&	0.30	\\
LiH	&	Li	&	7.43	&	-0.04	&	-0.03	&	-0.22	&		&	H	&	0.50	&	0.13	&	0.13	&	0.29	\\
	&	H	&	0.50	&	0.09	&	0.09	&	0.28	&		&	H	&	0.50	&	-0.14	&	-0.15	&	0.05	\\
LiF	&	Li	&	7.43	&	0.01	&	-0.06	&	-0.22	&	CO$_2$	&	C	&	37.68	&	-0.16	&	-0.33	&	-0.33	\\
	&	F	&	99.40	&	0.14	&	0.20	&	0.37	&		&	O	&	74.80	&	0.28	&	0.37	&	0.37	\\
NaCl	&	Na	&	161.87	&	0.09	&	-0.07	&	-0.20	&	CO	&	C	&	37.68	&	0.06	&	0.00	&	0.29	\\
	&	Cl	&	459.44	&	0.03	&	0.18	&	0.32	&		&	O	&	74.80	&	0.23	&	0.29	&	0.00	\\
HF	&	H	&	0.50	&	-0.19	&	-0.20	&	0.08	&	H$_2$O	&	O	&	74.80	&	0.48	&	0.50	&	0.12	\\
	&	F	&	99.40	&	0.34	&	0.35	&	0.08	&		&	H	&	0.50	&	-0.11	&	-0.13	&	0.06	\\
HCl	&	H	&	0.50	&	-0.03	&	-0.04	&	0.09	&	(H$_2$O)$_2$	&	O	&	74.80	&	0.48	&	0.51	&	0.13	\\
	&	Cl	&	459.44	&	0.17	&	0.17	&	0.04	&		&	H	&	0.50	&	-0.13	&	-0.14	&	0.06	\\
CH$_4$	&	C	&	37.68	&	0.26	&	0.21	&	-0.12	&		&	O	&	74.80	&	0.51	&	0.54	&	0.15	\\
	&	H	&	0.50	&	0.07	&	0.08	&	0.16	&		&	H	&	0.50	&	-0.11	&	-0.12	&	0.07	\\
CH$_3$CH$_3$	&	C	&	37.68	&	0.19	&	0.16	&	-0.09	&		&	H	&	0.50	&	-0.13	&	-0.14	&	0.05	\\
	&	H	&	0.50	&	0.09	&	0.09	&	0.18	&	H$_2$O$_2$	&	O	&	74.80	&	0.19	&	0.20	&	-0.03	\\
	&	H	&	0.50	&	0.09	&	0.09	&	0.18	&		&	H	&	0.50	&	-0.08	&	-0.10	&	0.14	\\
CH$_2$CH$_2$	&	C	&	37.68	&	0.21	&	0.19	&	0.02	&	NH$_3$	&	N	&	54.39	&	0.42	&	0.43	&	0.06	\\
	&	H	&	0.50	&	0.07	&	0.08	&	0.17	&		&	H	&	0.50	&	-0.03	&	-0.03	&	0.09	\\
CHCH	&	C	&	37.68	&	0.24	&	0.23	&	0.14	&	N$_2$H$_4$	&	N	&	54.39	&	0.21	&	0.23	&	-0.06	\\
	&	H	&	0.50	&	0.00	&	0.01	&	0.10	&		&	H	&	0.50	&	0.01	&	0.00	&	0.15	\\
CH$_3$OH	&	C	&	37.68	&	0.05	&	0.00	&	-0.13	&		&	H	&	0.50	&	0.00	&	-0.01	&	0.13	\\
	&	H	&	0.50	&	0.10	&	0.10	&	0.21	&		&		&		&		&		&		\\
	&	H	&	0.50	&	0.11	&	0.11	&	0.22	&		&		&		&		&		&		\\
	&	O	&	74.80	&	0.31	&	0.37	&	-0.04	&		&		&		&		&		&		\\
	&	H	&	0.50	&	-0.09	&	-0.10	&	0.11	&		&		&		&		&		&		\\
\hline
\end{tabular} 
\linespread{1.5}
\caption{Kinetic energy (t$_{A}$) of the molecular Hirshfeld-I fragments compared to the Hirshfeld-I fragmented molecular kinetic energies (t$_{A}^{h}$) and (t$_{A}^{h'}$). Values are presented relative to the kinetic energy (t$_{A}^{0}$) of the neutral isolated atom. Computations were performed at the ROHF/Aug-cc-pVTZ level of theory.}  
\label{kineticenergies}
\end{table} 
\linespread{1.5}

\subsection{Considerations about computational efficiency}

\begin{figure}[H]
\centering
\includegraphics[width=12cm] {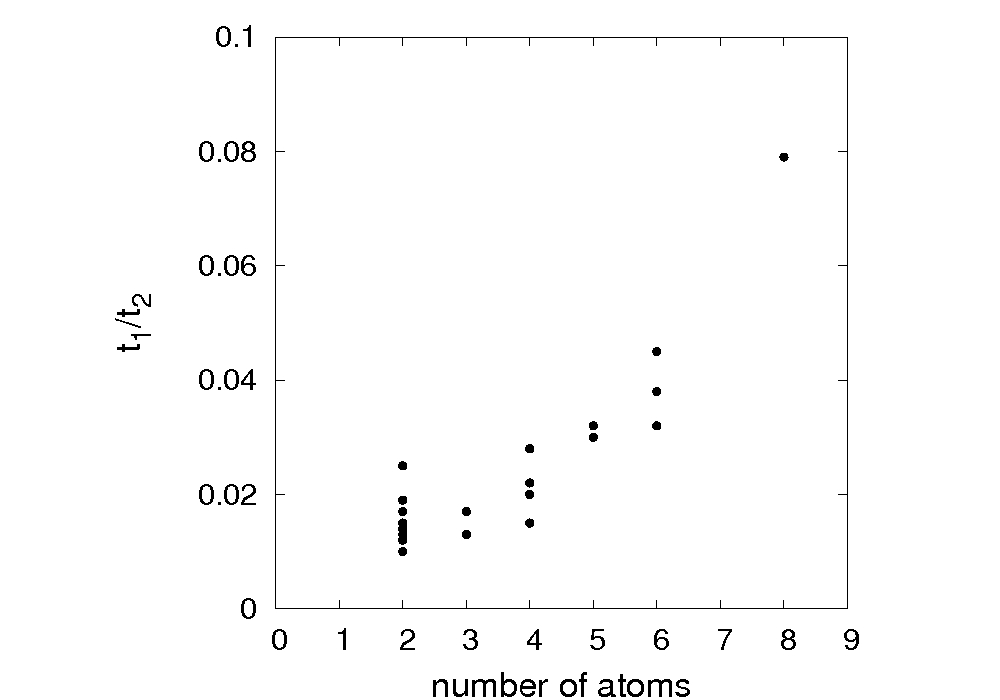}
\caption{\label{times} Calculation times for the energy decomposition (at the ROHF/Aug-cc-pVTZ level of theory) using the matrix approach ($t_{1}$) relative to the computational cost of using an r-space approach ($t_{2}$) as a function of the number of atoms in the molecule. 
}
\end{figure}

In this section, the computational efficiency of the approach to calculate atomic self-energies and the interaction energies in one-electron Hilbert-space (see Eqs. (\ref{selfenergy}) and (\ref{subinter}) ) is compared with the computational cost of the approach to compute these quantities using numerical integration in $\bm{r}$-space (see Eqs. (\ref{eqtenfrag}) and (\ref{fockinter})). When relatively small molecules and basis set sizes are considered, the matrix approach is computationally much less demanding than the $\bm{r}$-space approach. Figure \ref{times} displays the ratio of the computational costs of both approaches as a function of the number of atoms.  In order to obtain sufficiently accurate energy integrals (condensed to atoms and atom pairs) in the $\bm{r}$-space approach, one must go to very large grids, making the calculations quite time consuming. On the other hand, in the matrix approach, moderate grids suffice to construct the atom condensed atomic overlap matrices and the kinetic energy integrals over the molecular orbital basis are easily computed analytically.

Going to larger molecules and very large basis sets (e.g. Aug-cc-pVQZ), at one point the matrix approach will get computationally more demanding (at least at the Hartree Fock level) than the approach of Eq. (\ref{eqtenfrag}). In the former method all virtual molecular orbitals are included in the calculation, while in the latter method one only needs the occupied molecular orbitals. However, the crossover point is far from being reached in the present set of molecules.

\section{Conclusions}
\label{Conclusions}

In this paper, we applied a method for decomposing the one-electron molecular density matrix over the atoms in the molecule to calculate atom-condensed energy contributions at the Hartree-Fock level. 

In our opinion, this method is preferable to other approaches because it determines (Hartree-Fock) energy terms naturally associated with molecular fragments, whereas most other approaches fragment the molecular energy without assuring that there is an underlying electronic structure from which these energy fragments can be calculated explicitly. Since the Hartree-Fock energy can be expressed directly in terms of the molecular 1DM, energy terms are derived from the molecular 1DM fragments. For the current study, the 1DM fragments are consistent with the local and positive definite Hirshfeld-I partitioning of the electron density. We have shown that without this mathematical consistency, the results are necessarily ambiguous. 


Unlike in most cases where numerical integration is used intensively, the new scheme requires only the Hamiltonian matrix elements (one- and two-electron integrals) in Hilbert space, that are routinely computed in ab initio programs, and the atomic overlap integrals. 
Analysis of the computational efficiency of the new approach versus the more common one based on numerical integration of the molecular energy integrals weighted by atomic weight functions, shows that there is a trade-off between on the one hand the number of integrals that need to computed and the size of the grids required to reach an acceptable level of accuracy. When using numerical integration for, e.g., the kinetic energy, relatively fewer integrals need to computed but these are found to require large integration grids. When using the new density matrix formulation, more atomic overlap integrals are required but these can be computed with sufficient accuracy already for moderate grids. Analysis of the computational efficiency shows that the density matrix approach is computationally much more efficient.

\section{Acknowledgements}

We acknowledge support from Dr. Toon Verstraelen (CMM), who provided us with a Hirshfeld-I program. DVF, DVN, PB and DG acknowledge support from the research council (BOF) of Ghent University and FWO-Vlaanderen. PWA acknowledges support from Sharcnet, NSERC, and the Canada Research Chairs. 

\section{Appendix}
\label{Appendix}

When the molecular density matrix $\rho(\bm{r},\bm{r'})=\sum_{i}d_i \psi_i(\bm{r}) \psi_i(\bm{r'})$ is partitioned,
the atom-condensed kinetic energy based on $\rho_{A}(\bm{r},\bm{r'})$ can be written as:
\begin{eqnarray}
t_{A}&=&-\frac{1}{2}\int d\bm{r}d\bm{r'} \delta(\bm{r}-\bm{r'}) \left[ \nabla^2 \rho_A(\bm{r},\bm{r'}) \frac{}{} \right]  \nonumber \\
&=&-\frac{1}{2}\int d\bm{r}d\bm{r'} \delta(\bm{r}-\bm{r'}) \left[ \nabla^2 \rho(\bm{r},\bm{r'}) \frac{1}{2} [w_A(\bm{r})+w_A(\bm{r'})] \right] \nonumber \\
&=&-\frac{1}{2}\int d\bm{r}d\bm{r'} \delta(\bm{r}-\bm{r'}) \left[ \left( \nabla^2 \rho(\bm{r},\bm{r'}) \right) \cdot \frac{1}{2} [w_A(\bm{r})+w_A(\bm{r'})]  \right.  \nonumber \\ && \left.+   \rho(\bm{r},\bm{r'})  \cdot \frac{1}{2}\nabla^2 w_A(\bm{r}) +  \left( \nabla \rho(\bm{r},\bm{r'}) \right) \cdot \frac{}{}\nabla w_A(\bm{r}) \right] \nonumber \\ 
&=&
- \frac{1}{2} \sum_{i}d_i \int d{\bm{r}} \left[ \frac{}{} \psi_i(\bm{r}) \left( \nabla^{2} \psi_i(\bm{r}) \right) w_A(\bm{r}) \right. \nonumber \\ 
&& \left. + \left( \psi_i(\bm{r})\right)^2 \frac{1}{2} \left( \nabla^2 w_A(\bm{r}) \right) 
+ \psi_i(\bm{r}) \left( \nabla \psi_i(\bm{r}) \right) \cdot \nabla w_A(\bm{r})
 \right].
\end{eqnarray}
Removing $\nabla^2 w_A(\bm{r})$ and $\nabla w_A(\bm{r})$ using partial integration, Eqs. (\ref{eqtfrag}) and (\ref{eqtenfrag}) are shown to be equivalent:
\begin{eqnarray}
t_{A}&=&-\frac{1}{2} \sum_{i}d_i \int d{\bm{r}} w_A(\bm{r}) \left[ \psi_i(\bm{r})  \nabla^{2} \psi_i(\bm{r})  \frac{}{}\right. \nonumber \\ 
&& \left. + \frac{1}{2}  \nabla^2 \left( \psi_i(\bm{r})^2 \right)  - \nabla \left( \psi_i(\bm{r}) \cdot \nabla \psi_i(\bm{r}) \right) \right] \nonumber \\
&=&-\frac{1}{2} \sum_{i}d_i \int d{\bm{r}} w_A(\bm{r}) \left[ \psi_i(\bm{r}) \left( \nabla^{2} \psi_i(\bm{r}) \right) \frac{}{}\right] \nonumber \\
&=&
-\frac{1}{2} \int d\bm{r}d\bm{r'} \delta(\bm{r}-\bm{r'}) w_{A}(\bm{r}) \left[ \nabla^2 \rho(\bm{r},\bm{r'}) \frac{}{} \right] = t^h_A.
\end{eqnarray}

To demonstrate that the atom-condensed kinetic energy based on $\rho_{A}(\bm{r},\bm{r'})$ does not depend on the representation of the kinetic energy operator, we repeat the former exercise starting from a different respresentation of the kinetic energy operator:
\begin{eqnarray}
t_{A}&=&\frac{1}{2}\int d\bm{r}d\bm{r'} \delta(\bm{r}-\bm{r'}) \left[ \nabla \cdot \nabla' \rho_A(\bm{r},\bm{r'}) \frac{}{} \right]  \nonumber \\
&=&\frac{1}{2}\int d\bm{r}d\bm{r'} \delta(\bm{r}-\bm{r'}) \left[ \nabla \cdot \nabla' \rho(\bm{r},\bm{r'}) \frac{1}{2} [w_A(\bm{r})+w_A(\bm{r'})] \right] \nonumber \\
&=&\frac{1}{2}\int d\bm{r}d\bm{r'} \delta(\bm{r}-\bm{r'}) \left[ \left( \nabla \cdot \nabla' \rho(\bm{r},\bm{r'}) \right) \cdot \frac{1}{2} [w_A(\bm{r})+w_A(\bm{r'})]  \right.  \nonumber \\ && \left.+  \left( \nabla \rho(\bm{r},\bm{r'}) \right) \cdot \frac{1}{2}\nabla'w_A(\bm{r'}) +  \left( \nabla' \rho(\bm{r},\bm{r'}) \right) \cdot \frac{1}{2}\nabla w_A(\bm{r}) \right] \\ \nonumber
&=& \frac{1}{2} \sum_{i}d_i \int d{\bm{r}} \left[ \left( \nabla \psi_i(\bm{r}) \right)^{2} w_{A}(\bm{r}) 
+ \psi_i(\bm{r})\left( \nabla \psi_i(\bm{r}) \right) \cdot \nabla w_A(\bm{r})
\frac{}{}\right].
\end{eqnarray}
Removing $\nabla w_A(\bm{r})$ using partial integration, Eqs. (\ref{twokin}) and (\ref{eqtenfrag}) are seen to be equivalent as well:
\begin{eqnarray}
t_{A} &=&
\frac{1}{2} \sum_{i}d_i \int d{\bm{r}} w_A(\bm{r}) \left[ \left( \nabla \psi_i(\bm{r}) \right)^{2} - \nabla \cdot \left( \psi_i(\bm{r}) \nabla \psi_i(\bm{r}) \right) \right] \nonumber \\
&=&
-\frac{1}{2} \sum_{i}d_i \int d{\bm{r}} w_A(\bm{r}) \left[ \psi_i(\bm{r})  \nabla^2 \psi_i(\bm{r}) \right] \nonumber \\
&=&
-\frac{1}{2} \int d\bm{r}d\bm{r'} \delta(\bm{r}-\bm{r'}) w_{A}(\bm{r}) \left[ \nabla^2 \rho(\bm{r},\bm{r'}) \frac{}{} \right] = t^h_A.
\end{eqnarray}

\bibliographystyle{unsrt}

\end{document}